# Quantum Sensing in Two-Dimensional Materials

Xiao-Jie Wang[1], Yang-Yi Chen[1], Hong-Hua Fang[1*]

[1]State Key Laboratory of Precision Measurement Technology and Instruments, Department of Precision Instrument, Tsinghua University, Beijing, 100084 China

Authors to whom correspondence should be addressed: *hfang@mail.tsinghua.edu.cn*

**Abstract**

Quantum-enhanced sensing exploits the coherent dynamics of two-level systems (TLSs) to achieve exceptional sensitivities and measurement precision that surpass classical detection limits. While platforms such as nitrogen-vacancy centers in diamond and rare-earth doped crystals have shown excellent performance, their integration with surfaces and external targets remains limited by bulk geometries. Two-dimensional (2D) van der Waals materials—particularly hexagonal boron nitride (hBN)—offer a compelling alternative, providing atomically thin hosts for spin defects with intrinsic surface proximity and environmental accessibility. These attributes enable high-resolution sensing of magnetic fields, strain, and temperature at the nanoscale. In this Perspective, we review recent progress in quantum sensing using spin defects in hBN, including the widely studied boron vacancy ($V_B^-$) and emerging carbon-related single-spin centers. We summarize protocols for spin initialization, coherent manipulation, and optical readout, and highlight demonstrated applications in hybrid architectures and extreme environments and discuss advances in deterministic defect engineering, coherence preservation at the 2D limit. Finally, we discuss future opportunities and challenges in realizing scalable, robust, and multifunctional quantum sensors based on 2D materials.

**Introduction**

Quantum-enhanced sensing leverages the coherence of two-level systems to detect external perturbations—such as magnetic fields, strain, electric fields, or temperature—with precision beyond classical limits.[1,2] By mapping these signals onto quantum phase or population changes, such sensors achieve exceptional sensitivity, spatial resolution, and non-invasive performance under extreme conditions.[3-5] Diverse platforms including ions, atoms, photonic systems, and solid-state spin defects have demonstrated broad applicability across condensed matter physics, materials science, and biology.[4-11]

Among them, solid-state defect-based systems have garnered particular attention for their robustness, scalability, and room-temperature operability.[12,13] Defect spins in diamond—most notably nitrogen-vacancy (NV) centers—have established themselves as a leading sensing platform, with demonstrated applications in nanoscale magnetic imaging, thermometry, and scanning probe microscopy.[2,14,15] Likewise, color centers in silicon carbide offer complementary advantages in terms of CMOS compatibility and telecom-range emission.[16-19] Despite their success, traditional three-dimensional host materials face inherent limitations in proximity-limited sensing applications,[20] thereby reducing their sensitivity to external fields originating from surfaces, interfaces, or low-dimensional materials. Furthermore, large-scale integration of these platforms with emerging nanophotonic and electronic devices remains technologically challenging.

Recently, hexagonal boron nitride (hBN),[21] a van der Waals material with atomic thickness and wide bandgap (~6 eV), has emerged as a promising host for quantum spin defects.[22-26] Its layered structure allows atomically close proximity between the defect center and the sensing target, making it especially advantageous for interface-sensitive measurements. Since the first observation of optically addressable spin defects in hBN in 2020,[22] rapid progress has been made in characterizing boron vacancies ($V_B^-$) and carbon-related spin centers, with demonstrations of room-temperature optically detected magnetic resonance (ODMR),[26-28] coherent spin manipulation,[29-31] and quantum sensing of magnetic fields, strain, and temperature.[32-37] The 2D nature of hBN also offers natural integration with hybrid material systems—such as ferromagnets, superconductors, and piezoelectrics—opening new pathways for proximity-enhanced and multifunctional sensing and towards chip-scale hybrid quantum systems.[38-40]

This perspective reviews the recent advances in quantum sensing using spin defects in hBN. We summarize basic physical properties of spin defect and quantum sensing protocols, including initialization, coherent control, and readout. We then discuss recent developments in spin defect engineering, coherence characterization, and deterministic creation. Finally, we discuss current challenges and emerging opportunities for realizing robust, scalable, and integrable quantum sensors in two-dimensional platforms.

## 2. Coherent control of spin defect in hBN

Most of quantum sensing protocols fundamentally rely on a sequence of operations applied to a two-level quantum system:[1] state initialization (via optical pumping or a calibrated π-pulse), controlled interaction with the external physical parameter of interest, coherent manipulation (e.g., through Rabi oscillations, Ramsey interferometry, or spin echo techniques), and final readout (using photoluminescence, optically detected magnetic resonance (ODMR), or quantum interference). The two-level architecture offers the minimal yet complete Hilbert space necessary to support quantum coherence, accumulate phase information under perturbation, and perform high-fidelity readout—key ingredients for achieving sensitivities that surpass classical limits.

In the following sections, we focus on two main categories: the negatively charged boron vacancy ($V_B^-$) centers, which have been extensively studied, and the emerging class of carbon-related single spin defect, which offer promising coherence and optical addressability. Both systems serve as platforms for implementing quantum sensing protocols in atomically thin materials.

**2.1 $V_B^-$ Spin Defects in hBN**

Since the discovery of optically active quantum defects in both monolayer and few-layer hBN in 2016,[41] the development of hBN-based quantum platforms has accelerated significantly.[42-48] Among the various spin defects reported in hBN, the $V_B^-$ center has emerged as the most extensively studied candidate for quantum sensing applications. In 2018, Toledo et al. identified paramagnetic defects in neutron-irradiated hBN via conventional electron paramagnetic resonance (EPR) spectroscopy.[49] In 2020, Gottscholl et al. observed ODMR from ensembles of $V_B^-$ centers,[22] characterized by a broad photoluminescence (PL) spectrum centered around 850 nm (Fig. 1a). Later theoretical investigations provided a more detailed picture of the electronic and atomic structure of the $V_B^-$ defect.[37,50,51] The $V_B^-$ defect is believed to consist of a missing boron atom adjacent to a localized negative charge, forming a stable spin-triplet ground state (S = 1). This configuration yields three spin sublevels, $|m_s = 0\rangle$ and $|m_s = \pm 1\rangle$, which are split even at zero magnetic field due to intrinsic dipolar spin-spin interactions. The simplified energy level diagram for the $V_B^-$ is shown in Fig. 1b.[52] The ground-state electron spin Hamiltonian $H_{gs}$ of $V_B^-$ involves terms with ZFS, electron Zeeman splitting, and electron-nuclear spin hyperfine interaction:[22,29,37]

$$H_{gs} = D_{gs}\left[S_z^2 - \frac{S(S+1)}{3}\right] + E_{gs}(S_x^2 - S_y^2) + \gamma_e \mathbf{B}\mathbf{S} + \sum_{k=1,2,3} \mathbf{S}\mathbf{A_k}\mathbf{I_k}$$

Where $D_{gs} \approx 3.48$ GHz is the ground state longitudinal ZFS parameter, and $E_{gs} \approx 48$ MHz is transverse ZFS parameter due to residual strain and local electric fields in the lattice, **S** and $\mathbf{S}_j$ (j = x, y, z) are electron spin-1 operators. The Zero-field splitting of excited state triplet is $D_{es}$

≈ 2.1 GHz for longitudinal splitting and $E_{es}$ ≈ 48 MHz for transverse splitting, respectively.[52-54] These sublevels can be split further in the presence of an external magnetic field due to the Zeeman effect. A representative ODMR spectra is shown in Fig. 1c.[52] When the external magnetic field $B_0$ is parallel to the $c$ axis, the resonant frequencies of transitions between the $|m_s=0\rangle$ and $|m_s=\pm1\rangle$ states are

$$v_{\pm} = D_{gs} \pm \sqrt{E_{gs}^2 + (\gamma_e B_0)^2}$$

The magnetic field detection sensitivity can be written as

$$\eta \approx \frac{\Delta v}{\gamma_e C \sqrt{I_0}}$$

where $\Delta v$ is the linewidth of the spectrum, $C$ is the ODMR contrast, and $I_0$ is the detected rate of the photons. Obviously, these parameters are highly correlated with the experimental conditions-such as optical excitation powers and microwave driving strength-which must be carefully balanced to optimize sensing performance and achieve maximum sensitivity. Regarding coherence properties, the spin-lattice relaxation time ($T_1$) of the $V_B^-$ center reaches approximately 18 μs at room temperature and extends to 12.5 ms at 20 K.[29] The pure dephasing time ($T_2^*$) is typically around 100 ns, while the spin coherence time ($T_2$) approaches ~2 μs at room-temperature, as discussed in detail below.

To deterministically generate such defects, a variety of approaches have been explored, including neutron irradiation,[22,55,56] ion implantation,[57,58] electron beam irradiation,[59] and femtosecond laser writing.[60] Among these, ion implantation stands out as the most controllable and reproducible method currently for creating $V_B^-$ centers by introducing vacancies through energetic ion bombardment. Implantation using ion species such as H$^+$, N$^+$, He$^+$, C$^+$, Ar$^+$, Ga$^+$ and Xe$^+$ has successfully yielded $V_B^-$ ensembles with controllable lateral positions and dose densities.[57,58,61] Femtosecond laser writing offers a maskless, flexible alternative and has been used to generate optically active defects in hBN,[60,62,63] and other wide-bandgap hosts such as diamond,[64,65] silicon carbide,[66] and aluminium nitride.[67,68] However, in hBN, laser-induced spin defects typically suffer from lattice damage, leading to poor coherence and unstable emission.[60] Recent advances in near-threshold laser writing have enabled the generation of quantum emitters with improved brightness, sharp zero-phonon lines, and enhanced photostability,[69] though coherent spin control of such defects remains a major challenge.

Despite these achievements, $V_B^-$ centers continue to exhibit key limitations for quantum sensing. Their PL emission is broad and suffers from low quantum efficiency, reducing photon collection rates and hindering single-defect addressability.[51,70] Moreover, to date, no unambiguous demonstration of a single, optically active $V_B^-$ spin defect under ambient conditions has been reported, restricting the use of $V_B^-$ centers in protocols requiring single-spin initialization and readout.

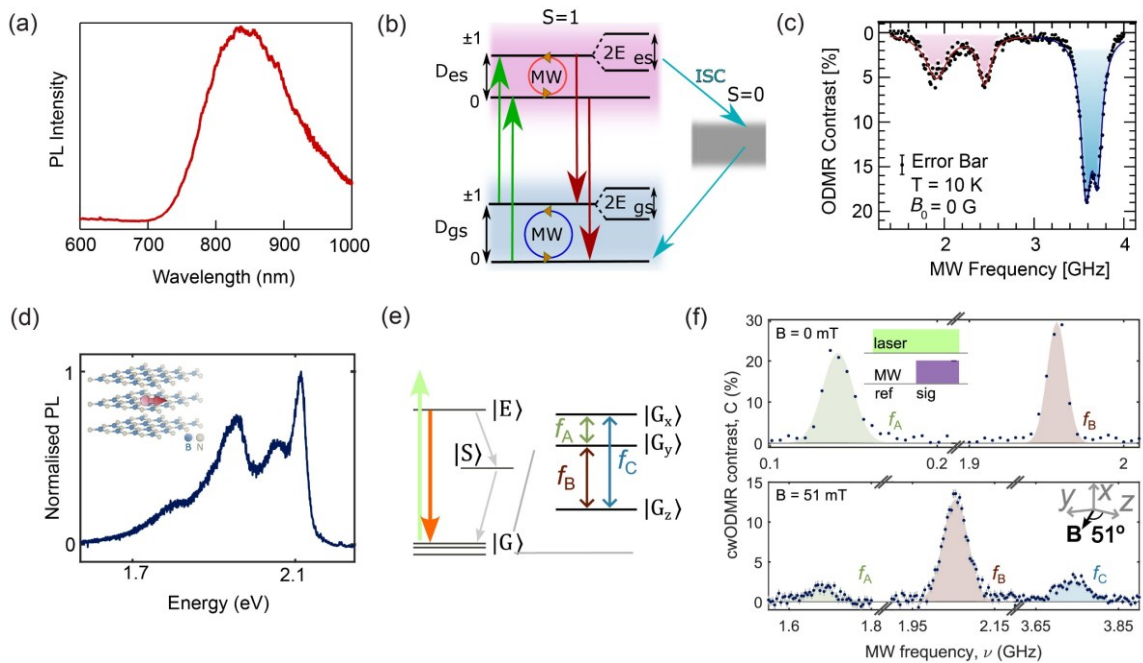

**Figure 1. Basic physical characteristics of representative spin defect in hBN.** (a) Room-temperature photoluminescence (PL) spectrum from an ensemble of $V_B^-$ defects, exhibiting a pronounced emission peak centered at ~850 nm. Reproduced with permission from Gottscholl et al. Nat. Mater. 19, 540–545 (2020). Copyright 2020, Nature Publishing Group.[22] (b)

Simplified energy level diagram of the $V_B^-$ center, showing a spin-triplet (S = 1) ground state (GS) and excited state (ES), along with a spin-singlet metastable state (MS). Reproduced with permission from Mathur et al. Nat. Comm. 13, 3222 (2022). Copyright 2022, Nature Publishing Group.[52] (c) Zero-field ODMR spectrum, displaying distinct resonance dips corresponding to spin transitions in the ES (orange) and GS (blue), with their respective zero-field splittings. Reproduced with permission from Mathur et al. Nat. Comm. 13, 3222 (2022). Copyright 2022, Nature Publishing Group.[52] (d) PL signal from carbon-related defects. (e) The diagram shows the electronic levels of carbon-related defects, with ground, excited, and metastable states. (f) The ODMR spectra at 0 mT and 51 mT show three spin transitions between the spin sublevels of an $S = 1$ system. (d-f) Reproduced with permission from Stern et al. Nat. Comm. 16, 4947 (2025). Copyright 2025, Nature Publishing Group.[33]

## 2.2 Single carbon-related spin defect

In parallel, a promising direction has emerged in the form of carbon-related spin defects, typically formed by substitutional or interstitial carbon impurities—exhibit both S = 1/2 and S = 1 spin configurations,[27,28,31] with better optical stability and narrower PL lines than $V_B^-$ ensembles at single-defect level.[30] More importantly, these defects show strong potential for coherent control at the single-spin level, a critical capability for scalable quantum sensing and information processing.

In 2021, Chejanovsky et al. firstly reported isolated spin emitters originating from substitutional impurities in hBN, exhibiting a longitudinal relaxation time $T_1$=17 μs and a dephasing time $T_2^*$=57 ns, thus supporting coherent microwave control via Rabi and Ramsey sequences.[26] In 2022, Stern et al. further observed ODMR spectra from individual carbon-related spin centers (Fig. 1d), corroborating their optical addressability and magnetic sensitivity.[27] In another study, a carbon-based defect exhibiting a triplet ground state was identified with a ODMR contrast above 30%,[31] significantly expanding the diversity of accessible spin manifolds in hBN. This defect demonstrated a magnetic field sensitivity of 3 μT Hz$^{-1/2}$, which is of the same order as that of the well-established NV center in diamond, and coherent spin control was confirmed via pulsed ODMR protocols. Furthermore, the S=1 carbon-related spin defect in hBN has been developed with a focus on its photodynamics, which contribute to its remarkable performance as a vectorial atomic sensor (Figs. 1d-f). This system operates efficiently under transverse magnetic field exceeding 150 mT and demonstrates DC sensitivity on the order of ~ µT Hz$^{-1/2}$. This combination of characteristics highlights the spin defect's sensing potential and its ability to serve as a highly precise vectorial sensor.[33] Most recently, a study from Purdue University further demonstrated atomic-scale nuclear magnetic resonance (NMR) using individual carbon-related defects.[71] The researchers achieved coherent control of nearby $^{13}$C nuclear spins with a π-gate fidelity exceeding 99.75% at room temperature. The measured nuclear spin coherence times-$T_2^*$=16.6 µs and $T_2$=162 µs—are orders of magnitude longer than those of the electron spin in hBN, underscoring the potential of these nuclear spins as long-lived quantum memories or registers. Collectively, these advances position single carbon-related spin defects as a highly promising platform for quantum-enhanced sensing in two-dimensional materials, offering enhanced optical readout contrast, spectral stability, and robust spin–nuclear coupling.

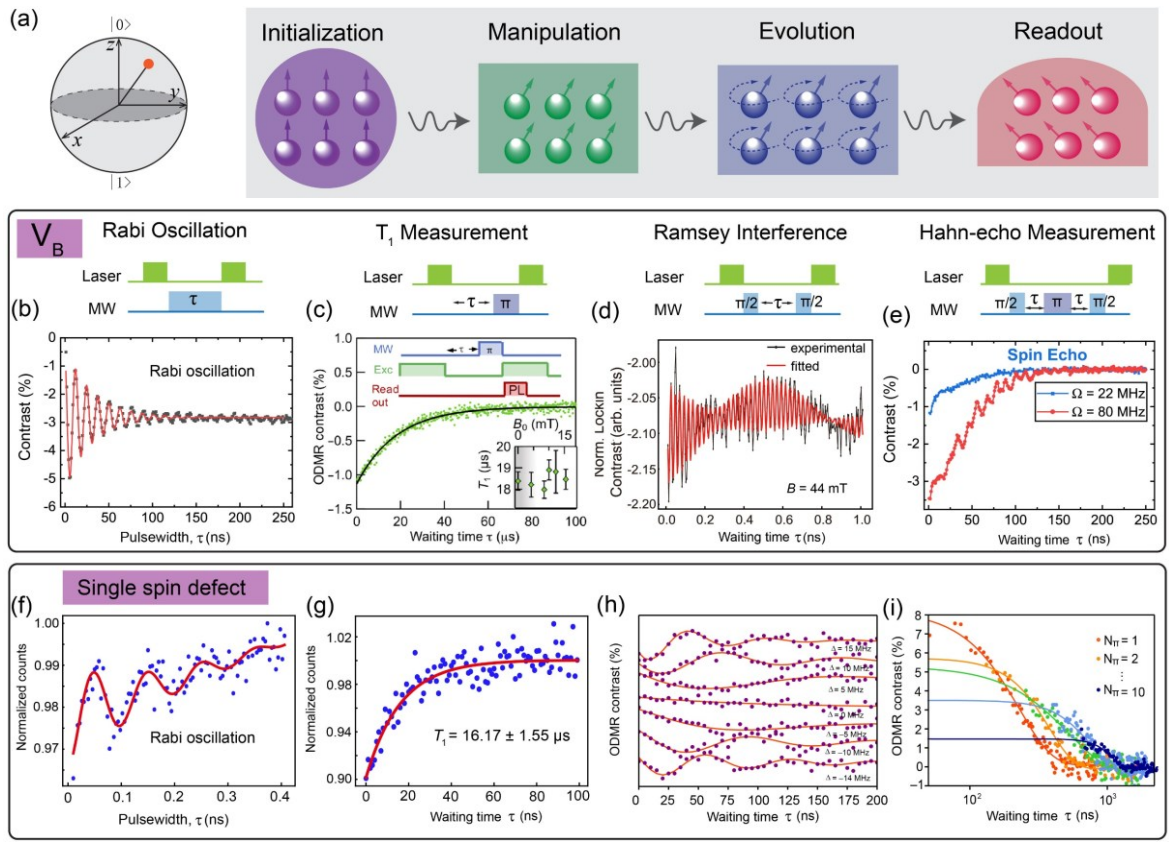

**Figure 2. Coherent spin control of quantum defects in hBN.** (a) Schematic overview of time-domain quantum sensing protocols, including initialization, coherent microwave control, quantum state evolution and optical readout. (b–e) Representative measurements of spin coherence in an ensemble of $V_B^-$ defects: (b) Rabi oscillations demonstrating coherent population transfer. Reproduced with permission from Ramsay et al. Nat. Comm. 14, 461 (2023). Copyright 2023, Nature Publishing Group.[72] (c) longitudinal spin relaxation ($T_1$) measurement, Reproduced with permission from Gottscholl et al. Sci. Adv. 7, eabf3630 (2021). Copyright 2021, licensed under a Creative Commons Attribution (CC BY) license.[29] (d) Ramsey interference revealing phase coherence and dephasing time ($T_2^*$) Reproduced with permission from Liu et al. Nat. Comm. 13, 5713 (2022). Copyright 2022, Nature Publishing Group.[73] and (e) Hahn-echo sequence extending spin coherence to $T_2$. Reproduced with permission from Ramsay et al. Nat. Comm. 14, 461 (2023). Copyright 2023, Nature Publishing Group.[72] (f–i) Analogous measurements performed on representative carbon-related single spin defects in hBN, illustrating coherent control at the single-spin level under ambient conditions. (f-g) Reproduced with permission from Guo et al. Nat. Comm. 14, 2893 (2023). Copyright 2023, Nature Publishing Group.[30] (h-i) Reproduced with permission from Stern et al. Nat. Mater. 23, 1379-1385 (2024). Copyright 2024, Nature Publishing Group.[31]

## 2.3 Coherent control of the spin defect.

To fully exploit the quantum sensing capabilities of spin defects in hBN, it is essential to access and manipulate their coherent dynamics within the ground-state manifold.[1,74] These spin defects—such as $V_B^-$ and carbon-related centers—offer optical initialization, microwave-driven manipulation, and spin-dependent photoluminescence readout, forming the basis for quantum sensing protocols (Fig. 2a). While continuous-wave (CW) ODMR provides basic spectroscopic access to spin transitions, its sensitivity is often limited by power broadening from laser and microwave excitation. To overcome these limitations, time-domain protocols such as pulsed ODMR, Ramsey interferometry, and spin echo have been developed and successfully adapted to hBN platforms. These methods rely on microwave-driven spin rotations and spin-dependent photoluminescence to extract information with sensitivity bounded by intrinsic coherence times ($T_2^*$, $T_2$) rather than power-broadened linewidths.[6,75]

Pulsed ODMR is a key method to overcome the limitations of CW-ODMR. In pulsed ODMR (Fig. 2b), the spin is first initialized to the $|m_s = 0\rangle$ state using a short laser pulse. After turning off the laser, a microwave pulse is applied to drive spin transitions without additional optical perturbation. A second laser pulse reads out the population through spin-dependent fluorescence. Varying the microwave duration reveals Rabi oscillations, which reflect coherent population transfer and allow extraction of the Rabi frequency ($\Omega_R$)—a crucial parameter for defining π- and π/2-pulse durations. Once π-pulses are calibrated, the spin-lattice relaxation time $T_1$ can be measured, with $T_1$ setting the upper bound for quantum coherence and indicating robustness to phonon or environmental noise—an important consideration for sensing in harsh conditions using hBN defects. This involves optical initialization, a variable dark interval (without optical or microwave fields), and a final π-pulse followed by readout (Fig. 2c). The resulting fluorescence decay over time τ reflects the thermal relaxation dynamics. The $T_1$ relaxation time of $V_B^-$ centers was shown to increase from 18 μs at room-temperature to 12.5 ms at 20 K, indicating strong temperature dependence.[29] The spin coherence time $T_2$ was measured to be 2 μs at room temperature under an applied magnetic field 8.5 mT.

Building on this control, Ramsey interferometry employs two π/2 pulses separated by a free evolution period τ to form a coherent superposition state sensitive to DC magnetic fields (Fig. 2d). During this interval, external fields induce phase accumulation, which is mapped into population contrast by the second π/2 pulse. The resulting Ramsey fringes oscillate with τ, allowing extraction of magnetic field strength with sensitivity limited by $T_2^*$—the inhomogeneous dephasing time. Gottscholl et al. measure the pure coherent time of $V_B^-$ is on the order of 100 ns.[29] To further extend coherence, the Hahn echo sequence introduces a π-pulse halfway through the free precession, reversing static phase errors and revealing the intrinsic $T_2$ (Fig. 2e), which reflect overall spin coherence time $T_2$ is on the order of ~2 μs. In hBN, $T_2$ is limited by the high concentration of nuclear spins in the lattice, which can be isolated from electronic spin use dynamical decoupling methods or reduce the concentration nuclear spins in the lattice. These protocols—demonstrated on both $V_B^-$ ensemble spin defects in hBN—enable high-fidelity field sensing with quantum-limited precision, making hBN a compelling platform for scalable, two-dimensional quantum metrology.

Significant progress has been achieved in the coherent control of single carbon-related spin defects in hexagonal boron nitride (hBN). Guo et al. reported Rabi oscillations and Hahn-echo interference patterns from isolated ultrabright carbon-related emitters, demonstrating robust quantum coherence at the single-spin level.[30] Their measurements revealed coherence times of $T_1 \approx 16$ μs, $T_2 \approx 2.5$ μs, and $T_2^* \approx 140$ ns, validating the feasibility of room-temperature quantum control (Fig. 2f and Fig. 2g). In a separate study, Stern et al. realized room-temperature coherent manipulation of an individually addressable single-photon emitter possessing a spin-triplet ground state with a zero-field splitting (ZFS) of 1.96 GHz.[31] The defect exhibited a dephasing time of $T_2^* \approx 100$ ns under the Ramsey interferometry (Fig. 2h) and a spin-echo coherence time of $T_2 \approx 200$ ns (Fig. 2i), confirming its applicability to time-domain quantum sensing protocols. Furthermore, Gao et al. demonstrated coherent control of a proximal $^{13}C$ nuclear spin in hBN, reporting nuclear spin coherence times of $T_2^* \approx 16.6$ μs and $T_2 \approx 162$ μs at room temperature—orders of magnitude longer than those of the electron spin. First-principles calculations suggest that donor–acceptor pair (DAP) complexes and $C_BO_N$ (carbon-boron-nitrogen-related complexes) are likely responsible for the observed spin centers, as they exhibit large hyperfine interactions consistent with experimental observations.

## 3. Applications of the spin defects in hBN

Rapid advances in nanoscale quantum sensing using hBN-based spin defects have been reported over the past few years. Owing to its atomic thickness and layered van der Waals structure, hexagonal boron nitride (hBN) offers exceptional spatial proximity between embedded spin defects and external samples. his ultrathin configuration enables strong, localized coupling to external fields, making hBN an attractive platform for hybrid quantum devices and interface-sensitive sensing applications. Building on this structural advantage, a growing array of quantum sensing modalities has been demonstrated using hBN spin defects, including magnetic field imaging, strain mapping, and nanoscale thermometry. A prominent example is the work by Healey et al., who assembled two distinct heterostructures composed of the van der Waals ferromagnet CrTe$_2$ and spin-defect-rich (V$_B^-$) hBN layers flake (Fig. 3a).[76] Their platform enabled time-resolved imaging of both local temperature and magnetic field dynamics near the Curie transition of the CrTe$_2$ layer, achieving sensitivities of approximately 2 mT Hz$^{-1/2}$ for magnetic fields and 80 mK Hz$^{-1/2}$ for temperature, respectively. Additionally, $T_1$-relaxometry-based sensing has been demonstrated under cryogenic conditions using wide-field V$_B^-$ ensembles.[77] In particular, one study visualized nanoscale magnetic phase transitions and spin fluctuations in Fe$_3$GeTe$_2$ (FGT), a prototypical 2D ferromagnet, achieving magnetic field sensitivity as high as 8 µT Hz$^{-1/2}$. These demonstrations highlight the ability of hBN-based quantum sensors to operate in spatially resolved wide-field modes and to interrogate phase-critical dynamics in low-dimensional materials.

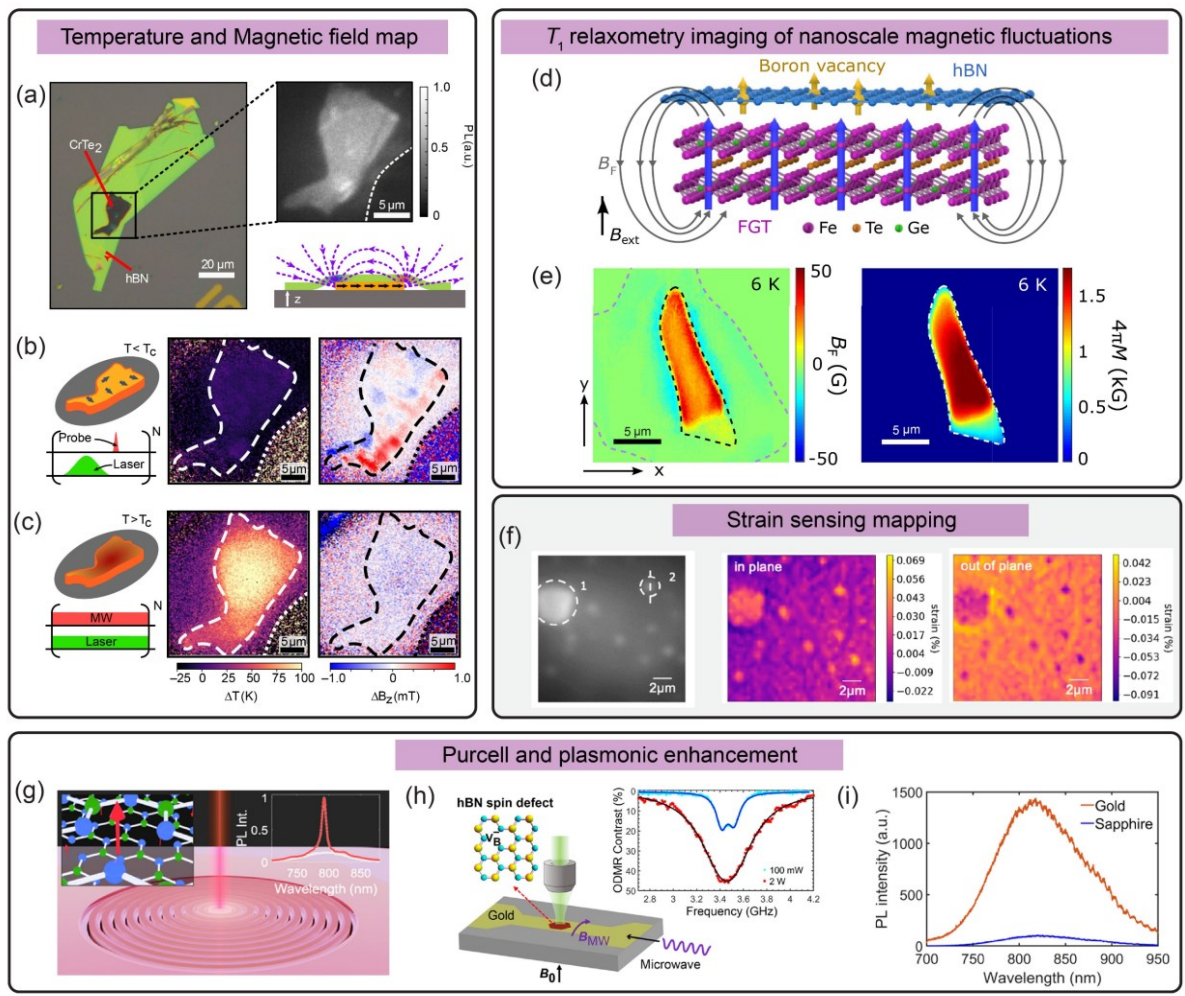

**Figure 3. Applications of quantum sensing based on spin defects in hBN and optical enhancement strategies.** (a) Schematic of a hybrid quantum sensing platform integrating $CrTe_2$ with spin-defect-enriched hBN, enabling simultaneous nanoscale magnetic and thermal imaging. (b–c) Pulse sequences used for time-resolved sensing (green: laser pulse; red: microwave pulse) and corresponding magnetization and temperature maps of the $CrTe_2$ flake. (a-c) Reproduced with permission from Healey et al. Nat. Phys. 19, 87-91 (2023). Copyright 2023, Nature Publishing Group.[76] (d) Schematic illustration of $T_1$ relaxometry used for imaging magnetic fluctuations in $Fe_3GeTe_2$ at cryogenic temperatures. (e) Two-dimensional maps of the static stray magnetic field and the reconstructed magnetization $4\pi M$ of an exfoliated $Fe_3GeTe_2$ flake, measured at 6 K under an applied perpendicular magnetic field **B**=142 G. Black and purple dashed lines indicate the boundaries of the $Fe_3GeTe_2$ and hBN flakes, respectively. (d-e) Reproduced with permission from Huang et al. Nat. Comm. 13, 5369 (2022). Copyright 2022, Nature Publishing Group.[77] (f) Strain mapping using $V_B^-$ defects in monolayer hBN with interfacial bubbles, showing spatially resolved ODMR frequency shifts. Reproduced with permission from Lyu et al. Nano Lett. 22, 6553-6559 (2022). Copyright 2022, American Chemical Society.[35] (g) Purcell enhancement of photoluminescence via integration of hBN spin defects into a bullseye cavity. Reproduced with permission from Froch et al. Nano Lett. 21, 6549-6555 (2021). Copyright 2021, American Chemical Society.[78] (h) Plasmonic enhancement using a gold-film microwave waveguide and result increased PL brightness (i) through localized surface plasmon coupling. (h-i) Reproduced with permission from Gao et al. Nano Lett. 21, 7708-7714 (2021). Copyright 2021, American Chemical Society.[79]

Strain sensing has also emerged as a key application of hBN-based quantum sensors. The $V_B^-$ center in hBN exhibits ODMR frequency shifts in response to local lattice distortions, enabling quantitative strain measurements at the nanoscale. Lyu et al. leveraged this property to image the strain in monolayer hBN containing interfacial bubbles, revealing both in-plane and out-of-plane strain components.[35] By correlating ODMR shifts with Raman spectroscopy, they extracted strain magnitudes ranging from 0.01% to 0.1%. Similarly, Yang et al. applied $V_B^-$ to characterize the strain profile in hBN patterned over nanofabricated pillars, detecting local strain levels up to 0.015%.[36] Taken together, these results demonstrate that spin defects in hBN enable multifunctional, high-resolution sensing of diverse physical parameters—including magnetic fields, temperature, and lattice strain—with a spatial footprint defined by the sub-nanometer scale of the host lattice. Their compatibility with ambient conditions and integration into layered heterostructures further establishes hBN as a uniquely versatile platform for quantum-enhanced metrology in two-dimensional materials.

Despite the promising prospects of $V_B^-$ spin defects in hBN for quantum sensing, their practical utility remains limited by intrinsically low PL brightness, which result in the absence of confirmed single-spin optical readout. For high-sensitivity quantum sensing, maximizing the photon emission rate and improving photon collection efficiency are critical steps toward reliable signal readout. To address these limitations, one notable approach was demonstrated by Froch et al., who embedded hBN emitters into a bullseye cavity structure, achieving a 6.5-fold increase in emission intensity due to the enhanced light-matter interaction within the cavity mode (Fig. 3g).[78] This cavity provided both improved out-coupling efficiency and a modest Purcell factor through reduced mode volume. Besides, Gao et al. employed localized surface plasmons supported by a gold-film microwave waveguide to simultaneously enhance both the optical and spin properties of hBN spin defects (Fig. 3h).[79] The nanoscale surface roughness of the gold film supported strong plasmonic resonances, which led to a 17-fold enhancement in PL intensity (Fig. 3i). Beyond optical enhancement, the gold waveguide also acted as a microwave antenna, producing a strong in-plane microwave magnetic field that efficiently drove spin transitions. This dual-function platform enabled high-contrast optically detected magnetic resonance (ODMR) measurements at room temperature, significantly improving the signal-to-noise ratio and overall magnetic field sensitivity of the system. Beyond these, various other Purcell- or plasmonic-enhanced platforms have demonstrated significant improvements in photon emission and optical readout contrast for spin defects in hBN.[80-83]

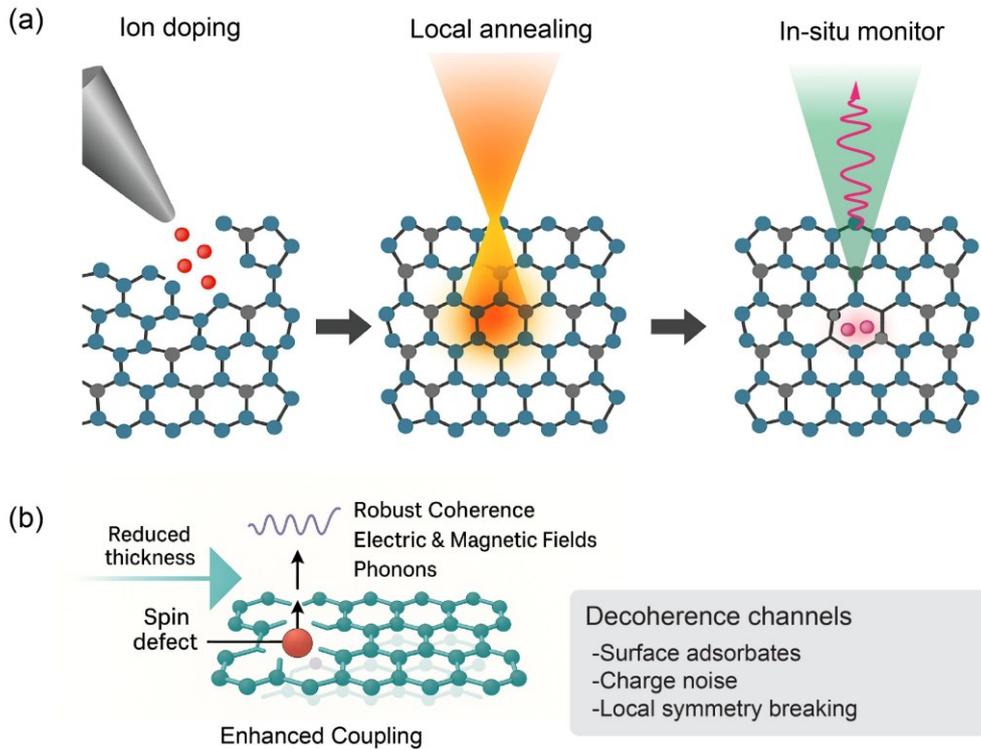

**Figure 4**: **Challenges and outlook for achieving robust quantum sensors.** (a) Schematic illustration of the fabrication workflow involving ion implantation followed by femtosecond laser annealing. Ion implantation introduces selected impurity species to form color centers, while spatially localized laser annealing enables precise defect activation and lattice healing. In-situ monitoring during the annealing process provides real-time feedback on defect formation dynamics, thereby enhancing the yield and reproducibility of on-demand spin centres. (b) Schematic representation of the challenges and opportunities associated with spin sensors in the monolayer limit. As material thickness approaches the atomic scale, spin defects experience increased environmental coupling and emerging decoherence pathways. These effects pose challenges to quantum coherence but also enable enhanced sensitivity for probing local magnetic, electric, and strain fields, opening new avenues for high-resolution quantum sensing at the nanoscale.

## 4. Discuss and Outlook

Although significant progress has been made in the field of quantum sensing based on spin defects in hBN, several critical challenges still remain. The performance of quantum sensors based on spin defects in hBN—critically depends on the deterministic creation of isolated, optically addressable quantum emitters with stable electronic and spin coherence properties. Notably, promising defects such as carbon-related spin centers have thus far only been observed in grown samples, limiting their practical integration into device platforms.

Controlled on-site doping plays a pivotal role in generating these quantum defects, as most spin-active centers arise from impurity configurations—such as substitutional carbon or nitrogen atoms, or vacancy–impurity complexes—that introduce unpaired spins and zero-field splitting. However, conventional stochastic methods like high-energy ion implantation or neutron irradiation lack spatial precision and reproducibility, limiting scalability and device-level integration. To overcome these challenges, recent efforts have turned toward site-controlled and feedback-guided fabrication (Fig. 4a). For example, Cheng et al. demonstrating a two-step fabrication method for tin vacancy centres in diamond that uses site-controlled ion implantation following by local femtosecond laser annealing with in-situ spectral monitoring, which also can monitor the transition between different defect state during defect informing process.[84] Additionally, near-threshold femtosecond laser writing has emerged as a promising technique,[64,67-69] capable of generating quantum emitters with minimal structural damage, preserving the surrounding lattice integrity and coherence time. Integrating in situ spectroscopy—such as confocal photoluminescence (PL), cathodoluminescence (CL), or photostability mapping—into the fabrication workflow provides real-time feedback,[65] enabling "write-and-verify" capabilities in a single cycle. This closed-loop methodology significantly improves defect yield, placement accuracy, and uniformity, facilitating scalable defect arrays for quantum devices. Moreover, the synergistic integration of diverse fabrication and characterization techniques not only improves engineering control but also inspires new insights into the fundamental defect formation dynamics—potentially unveiling novel pathways for defect generation and stabilization.

Another key challenge is the stabilization and control of defect charge states, which directly affects optical visibility and spin readout fidelity. Techniques such as local electrostatic gating, substrate engineering, and photoionization are actively being explored to maintain desirable charge states (e.g., $V_B^-$ in hBN) and to dynamically tune optical and spin properties for enhanced sensing performance.[85-87] Finally, strain and dielectric engineering offer powerful post-fabrication tools to tailor defect environments. By patterning the substrate or encapsulating layers, one can modulate strain fields, local dielectric constants, and phonon interactions—enabling tunability in emission wavelengths, zero-field splitting, and spin coherence properties. This paves the way for environment-responsive quantum sensors tailored to specific modalities such as pressure sensing, electric field detection, or nanoscale magnetometry.

As two-dimensional materials approach the atomic thickness limit, the physical behavior of embedded spin defects undergoes fundamental transformations (Fig. 4b).[88] The lack of bulk dielectric screening and the dominance of surface effects significantly enhance the sensitivity of defect spins to local electric, magnetic, and strain fields, making defects in hBN ideal candidates for nanoscale quantum sensing. However, this reduced dimensionality also alters coherence properties by modifying the phonon environment, zero-field splitting (ZFS), and spin–orbit interactions.[89-91] At the same time, new decoherence channels emerge due to surface adsorbates, charge noise, and local symmetry breaking. To fully realize the potential of these atomic-scale quantum sensors, further research is needed to elucidate the underlying defect physics and optimize their coherence and sensing performance at the single-atom level.

Overall, spin defects in hBN present a powerful platform for probing interfacial physics, catalytic reactions, and proximity-induced quantum phases—particularly when integrated into van der Waals heterostructures with ferromagnets, superconductors, or piezo electrics.[38,40] The atomically thin and mechanically flexible nature of 2D materials further facilitates seamless integration with on-chip photonic, radiofrequency (RF), and microelectromechanical (MEMS) systems, laying the groundwork for scalable, ultra-compact quantum sensors. Moving forward, a systematic research effort that combines deterministic defect creation, precision quantum control, and rigorous theoretical modelling will be essential to harness the sensing potential of 2D materials and realize their impact in next-generation quantum technologies.

**Acknowledgment:** The authors acknowledge the financial support from the National Natural Science Foundation of China (No. 62075115, 62335013) and the National Key R&D Program of China (No. 2022YFB4600400).

**Disclosures:** The authors declare no conflicts of interest.

**Data availability:** Data underlying the results presented in this paper are not publicly available at this time but may be obtained from the authors upon reasonable request.